%% file: root2.tex
\documentclass[conference,letterpaper]{IEEEtran}
\usepackage[utf8]{inputenc}
\usepackage{fancyhdr}
\usepackage{amsmath}%
\usepackage{graphicx}

\usepackage{cancel}

\usepackage{amsmath} 
\usepackage{amssymb} 
\usepackage{pifont} 
 \usepackage{cite}
\usepackage{url}
\usepackage{comment}
\usepackage{color, colortbl}
\usepackage{enumerate}
\usepackage{arcs}

\usepackage{fancyhdr}

\usepackage{pdfpages}

\newtheorem{defini}{Definition}
\newtheorem{thm}{Theorem}

\definecolor{Gray}{gray}{0.9}
\begin{document}


\title{{Witness-Functions versus Interpretation-Functions for Secrecy in Cryptographic Protocols: What to Choose?}}


\author{\IEEEauthorblockN{Jaouhar Fattahi$^1$, Mohamed Mejri$^1$, Marwa Ziadia$^1$, Takwa Omrani$^2$ and Emil Pricop$^3$}
\IEEEauthorblockA{$^1$Department of Computer  Science and Software  Engineering. Université Laval. Québec. Canada.}
\IEEEauthorblockA{$^2$Computer Science and Software Engineering Department. MACS Research Centre. University of Gabès. Tunisia.}
\IEEEauthorblockA{$^3$Automatic Control, Computers and Electronics Department Petroleum-Gas University of Ploiesti. Romania.}

}


%


\maketitle
\thispagestyle{plain}

\fancypagestyle{plain}{
\fancyhf{}	
\fancyfoot[L]{}
\fancyfoot[C]{}
\fancyfoot[R]{}
\renewcommand{\headrulewidth}{0pt}
\renewcommand{\footrulewidth}{0pt}
}

\pagestyle{fancy}{
\fancyhf{}
\fancyfoot[R]{}}
\renewcommand{\headrulewidth}{0pt}
\renewcommand{\footrulewidth}{0pt}

\begin{abstract}
Proving that a cryptographic protocol is correct for secrecy is a hard task. One of the strongest strategies to reach this goal is to show that it is increasing, which means that the security level of every single atomic message exchanged in the protocol, safely evaluated, never deceases. Recently, two families of functions have been proposed to measure the security level of atomic messages. The first one is the family of interpretation-functions. The second is the family of witness-functions. In this paper, we show that the witness-functions are more efficient than interpretation-functions. We give a detailed analysis of an ad-hoc protocol on which the witness-functions succeed in proving its correctness for secrecy while the interpretation-functions fail to do so.\\                                                                                                                                                                                                                                                                                                                                                                                                                                                                                                                                                                                                                                                                                                                                                                                                                                                                                                                                                                                                                                                                                                                                                                                                                                                                                                                                                                                                                                                                                                                                                                                                                                                                                                                                                                                                                                                                                                                                                                                                                                                                                                                                                                                                                                                                                                                                                                                                                                                                                                                                                                                                                                                                                                                                                                                                                                                                                                                                                                                                                                                                                                                                                                                                                                                                                                                                                                                                                                                                                                                                                                                                                                                                                                                                                                                                                                                                                                                                                                                                                                                                                                                                                                                                                                                                                                                                                                                                                                                                                                                                                                                                                                                                                                                                                                                                                                                                                                                                                                                                                                                                                                                                                                                                                                                                                                                                                                                                                                                                                                                                                                                                                                                                                                                                                                                                                                                                                                                                                                                                                                                                                                                                                                                                                                                                                                                                                                                                                                                                                                                                                                                                                                                                                                                                                                                                                                                                                                                                                                                                                                                                                                                                                                                                                                                                                                                                                                                                                                                                                                                                                                                                                                                                                                                                                                                                                                                                                                                                                                                                                                                                                                                                                                                                                                                                                                                                                                                                                                                                                                                                                                                                                                                                                                                                                                                                                                                                                                                                                                                                                                                                                                                                                                                                                                                                                                                                                                                                                                                                                                                                                                                                                                                                                                                                                                                                                                                                                                                                                                                                                                                                                                                                                                                                                                                                                                                                                                                                                                                                                                                                                                                                                                                                                                                                                                                                                                                                                                                                                                                                                                                                                                                                                                                                                                                                                                                                                                                                                                                                                                                                                                                                                                                                                                                                                                                                                                                                                                                                                                                                                                                                                                                                                                                                                                                                                                                                                                                                                                                                                                                                                                                                                                                                                                                                                                                                                                                                                                                                                                                                                                                                                                                                                                                                                                                                                                                                                                                                                                                                                                                                                                                                                                                                                                                                                                                                                                                                                                                                                                                                                                                                                                                                                                                                                                                                                                                                                                                                                                                                                                                                                                                                                                                                                                                                                                                                                                                                                                                                                                                                                                                                                                                                                                                                                                                                                                                                                                                                                                                                                                                                                                                                                                                                                                                                                                                                                                                                                                                                                                                                                                                                                                                                                                                                                                                                                                                                                                                                                                                                                                                                                                                                                                                                                                                                                                                                                                                                                                                                                                                                                                                                                                                                                                                                                                                                                                                                                                                                                                                                                                                                                                                                                                                                                                                                                                                                                                                                                                                                                                                                                                                                                                                                                                                                                                                                                                                                                                                                                                                                                                                                                                                                                                                                                                                                                                                                                                                                                                                                                                                                                                                                                                                                                                                                                                                                                                                                                                                                                                                                                                                                                                                                                                                                                                                                                                                                                                                                                                                                                                                                                                                                                                                                                                                                                                                                                                                                                                                                                                                                                                                                                                                                                                                                                                                                                                                                                                                                                                                                                                                                                                                                                                                                                                                                                                                                                                                                                                                                                                                                                                                                                                                                                                                                                                                                                                                                                                                                                                                                                                                                                                                                                                                                                                                                                                                                                                                                                                                                                                                                                                                                                                                                                                                                                                                                                                                                                                                                                                                                                                                                                                                                                                                                                                                                                                                                                                                                                                                                                                                                                                                                                                                                                                                                                                                                                                                                                                                                                                                                              
\end{abstract}


%
\IEEEpeerreviewmaketitle

\section{Introduction}

A cryptographic protocol is an encrypted communication between at least two agents. This messages' exchange is governed by a set of rules dictated by the protocol. The purpose of designing protocols using cryptography is to ensure the security of this communication, since it is always supposed to be executed in a hostile environment. In such insecure network, cryptography offers a robust set of techniques which counteract malicious intents and protect legitimate users. However, relying only on cryptography to achieve security goals is essential, it is insufficient. The most prominent example of this is the Needham-Schroeder protocol that had been deemed to be a secure protocol until it was paralyzed by a man-in-the-middle attack seventeen years after its first utilization. Hence, the need for analytical methods of cryptographic protocols' verification is widely accepted. Several methods have been developed and varied models for their verification have been proposed as well. Although promising efforts, the obtained results remains mixed, and this is plausible since that proving the correctness of cryptographic protocols is generally an undecidable problem ~\cite{cortier1971,ComonLundhKKKKK,Lundh1}. In this paper, we focus on two recent static methods, namely, the interpretation-functions and witness-functions. Both transform the  correctness problem (regarding the secrecy property) into a problem of protocol growth. They emit sufficient conditions on the metrics, that measure the security level of each atomic message exchanged in the protocol, and on the protocol itself, to deduce that it is an increasing protocol and consequently could be declared correct. Here, we show that the witness-functions are more efficient than the interpretation-functions in proving secrecy in cryptographic protocols. The analysis takes place in a role-based specification~\cite{Debbabi11}. We firstly extract the roles of each agent in the protocol. Then from these roles, we define the generalized roles according to the knowledge of each agent, while unknown messages are substituted by variables. In both analyses, we adopt the Dolev and Yao~\cite{DolevY1} conditions. We suppose that the intruder has the total control over the network: intercepting, blocking, forging and redirecting messages, anything except decrypting the message without possessing the decryption key. All the notations used in this paper are given in \cite{WitnessArt1}. We refer the reader to carefully take note of them before further reading.

\section{Paper organization}
This paper is organized as follows:
\begin{enumerate}
\item In Section \ref{sectionPreuveThFond}                                                                                                                                                                                                                                                                                                                                                                                                                                                                                                                                                                                                                                                                                                                                                                                                                                                                                                                                                                                                                                                                                                                                                                                                                                                                                                                                                                                                                                                                                                                                                                                                                                                                                                                                                                                                                                                                                                                                                                                                                                                                                                                                                                                                                                                                                                                                                                                                                                                                                                                                                                                                                                                                                                                                                                                                                                                                                                                                                                                                                                                                                                                                                                                                                                                                                                                                                                                                                                                                                                                                                                                                                                                                                                                                                                                                                                                                                                                                                                                                                                                                                                                                                                                       , we remind an important result: increasing protocols are correct for secrecy. We exhibit few conditions that must be satisfied by the functions used to evaluate the level of security of atomic messages;
\item In Section \ref{sectionWF}, we present the interpretation-functions~\cite{Houmani1,Houmani3,Houmani8,Houmani5} , their way of evaluating atomic messages and their way of treating variables;
\item In Section \ref{sectionWF}, we present the witness-functions~\cite{TheseJF,FattouhFattahiMGP16,TrustCom7301205,TrustComFattahiMP15,FattouhFattahiMP16,ChapterJFSpringer,WitnessArt1,Relaxed}, their way of evaluating atomic messages and their way of treating variables. We highlight their lower bounds and upper bounds and their usage of derivation to reduce variable impact;
\item In Section \ref{sectionAnWL2}, we analyze an ad-hoc protocol by both interpretation-functions and witness-functions. We show that it could be proven correct by the witness-functions but not by the interpretation-functions. In Section we explain why that has happened and we explain why witness-functions are more precise than  interpretation-functions;
\item In Section \ref{conc}, we conclude and we introduce to the future avenues of our research.

\end{enumerate}

\input{rappelSomeConfSMC2017WFvsIF}

\input{someProtocolSMC2017WFvsIF}

\section{Results, interpretation and related works}

As a result of our analysis, one witness-function succeeds in demonstrating the correctness of the protocol $p$ for secrecy, whereas the two interpretation-functions fail to do so. Indeed, the witness-functions have two major advantages compared to the interpretation-functions. On the one hand, the incorporated derivation in the upper bound of a witness-function precludes variables from playing any role in the received messages. On the other hand, variables in sent messages are carefully rummaged by the lower bound. These variables sometimes contain odd identities, interpreted as an intrusion, but often they do not, as it was the case in our protocol. This tightly depends on the protocol structure. As for the interpretation-functions, they treat variables with naivety. They just  return their content (i.e. agent identities) with no further inspection. This may often constitute an obstacle in showing the protocol growth, then, its correctness. A classical scenario on which the interpretation-function $\mbox{DEKAN}$ always fails is when an atomic secret $\alpha$ is received with a variable $X$ as a neighbor, and sent with another variable $Y$ as a neighbor. In that case, there is no hope to prove the growth of the protocol. Besides, in case of an analysis failure with a witness-function, we can figure out which encryption pattern causes this failure, and hence, we understand where and why a potential flaw may occur, if any. This indicates us with precision which item we have to modify in the protocol structure so that it becomes increasing, as well. The interpretation-functions do not have this capability to precisely explain why a flaw arises. They just give a raw and fuzzy indication about it. This indication is, in several cases, not helpful or even misleading. According to our experience, the false positive ratio related to interpretation-functions is much higher than the one related to the witness-functions. However, this precision has its cost. On the one hand, the protocol analysis with a witness-function is slower than the analysis with an interpretation-function. On the other hand, analyzing a protocol with a witness-function is sensitive to multi-protocol environments. In contrast, analyzing it with an interpretation-function is not, as these latter are universal.

\section{Conclusion and future work}\label{conc}

In this paper, we have exhibited a comparative study between witness-functions  and interpretation-functions throughout a detailed analysis of an ad hoc protocol. We managed to show that  the witness-functions may very much succeed where the interpretation-functions failed. In a future work, we will provide the formal proof that the witness-functions mask the interpretation-functions. That means, we will formally prove that, when the witness-functions fail, the interpretation-functions necessarily fail, and when the  interpretation-functions succeed, the witness-functions necessarily succeed. This will be the final step toward the deprecation of the interpretation-functions.



\bibliographystyle{ieeetr}
\bibliography{Ma_these} 


\section*{Notice}
\small{
© 2017 IEEE. Personal use of this material is permitted. Permission from IEEE must be obtained for all other uses, in any current or future media, including reprinting/republishing this material for advertising or promotional purposes, creating new collective works, for resale or redistribution to servers or lists, or reuse of any copyrighted component of this work in other works.
}
\end{document}

%% file: rappelSomeConfSMC2017WFvsIF.tex
\section{Correctness of Increasing Protocols}\label{sectionPreuveThFond}                                                                                                                                                                                                                                                                                                                                                                                                                                                                                                                                                                                                                                                                                                                                                                                                                                                                                                                                                                                                                                                                                                                                                                                                                                                                                                                                                                                                                                                                                                                                                                                                                                                                                                                                                                                                                                                                                                                                                                                                                                                                                                                                                                                                                                                                                                                                                                                                                                                                                                                                                                                                                                                                                                                                                                                                                                                                                                                                                                                                                                                                                                                                                                                                                                                                                                                                                                                                                                                                                                                                                                                                                                                                                                                                                                                                                                                                                                                                                                                                                                                                                                                                                                                                                                                                                                                                                                                                                                                                                                                                                                                                                                                                                                                                                                                                                                                                                                                                                                                                                                                                                                                                                                                                                                                                                                                                                                                                                                                                                                                                                                                                                                                                                                                                                                                                                                                                                                                                                                                                                                                                                                                                                                                                                                                                                                                                                                                                                                                                                                                                                                                                                                                                                                                                                                                                                                                                                                                                                                                                                                                                                                                                                                                                                                                                                                                                                                                                                                                                                                                                                                                                                                                                                                                                                                                                                                                                                                                                                                                                                                                                                                                                                                                                                                                                                                                                                                                                                                                                                                                                                                                                                                                                                                                                                                                                                                                                                                                                                                                                                                                                                                                                                                                                                                                                                                                                                                                                                                                                                                                                                                                                                                                                                                                                                                                                                                                                                                                                                                                                                                                                                                                                                                                                                                                                                                                                                                                                                                                                                                                                                                                                                                                                                                                                                                                                                                                                                                                                                                                                                                                                                                                                                                                                                                                                                                                                                                                                                                                                                                                                                                                                                                                                                                                                                                                                                                                                                                                                                                                                                                                                                                                                                                                                                                                                                                                                                                                                                                                                                                                                                                                                                                                                                                                                                                                                                                                                                                                                                                                                                                                                                                                                                                                                                                                                                                                                                                                                                                                                                                                                                                                                                                                                                                                                                                                                                                                                                                                                                                                                                                                                                                                                                                                                                                                                                                                                                                                                                                                                                                                                                                                                                                                                                                                                                                                                                                                                                                                                                                                                                                                                                                                                                                                                                                                                                                                                                                                                                                                                                                                                                                                                                                                                                                                                                                                                                                                                                                                                                                                                                                                                                                                                                                                                                                                                                                                                                                                                                                                                                                                                                                                                                                                                                                                                                                                                                                                                                                                                                                                                                                                                                                                                                                                                                                                                                                                                                                                                                                                                                                                                                                                                                                                                                       

We remind here a crucial result: ''A protocol is correct for secrecy when we analyze it with a safe function and we show that it is increasing''.

\subsection{Safe Functions}
\begin{defini}{(Well-Formed Function)}\label{bienforme}
{

\begin{center}
\begin{tabular}{lll}
   ${F}(\alpha,\{\alpha\})$ & $=$ & $\bot$ \\
   ${F}(\alpha, {M}_1 \cup {M}_2)$ & $=$ & ${F}(\alpha, {M}_1)\sqcap{F}(\alpha,{M}_2)$ \\
   ${F}(\alpha,{M})$ & $=$ & $\top, \mbox{ if } \alpha \notin {\mathcal{A}}({M})$ \\
\end{tabular}
\end{center}
}
\end{defini}

\begin{defini}{(Full-invariant-by-intruder Function)}\label{spi}
{$ $\\
$
 {M} \models_{\mathcal{C}} m \Rightarrow ({F}(\alpha,m) \sqsupseteq{F}(\alpha,{M})) \vee (\ulcorner K(I) \urcorner \sqsupseteq \ulcorner \alpha \urcorner).
$
}
\end{defini}

A function ${F}$ is safe iff it is well-formed and full-invariant-by-intruder.\\

\begin{defini}{(${F}$-Increasing Protocol)}\label{ProAbsCroi}
{
A protocol $p$ is ${F}$-increasing iff: $\forall R.r, {F}(\alpha, r^+\sigma)\sqsupseteq \ulcorner \alpha \urcorner \sqcap{F}(\alpha, R^-\sigma)$
}
\end{defini}

\begin{thm}{(Secrecy of Increasing Protocols)}\label{mainTh}
{
If ${F}$ is a safe function and $p$ is an ${F}$-increasing protocol then $p$ is correct for secrecy.
}
\end{thm}

\section{Interpretation-Functions}\label{sectionWF}

Houmani et al.~\cite{Houmani1,Houmani3,Houmani8,Houmani5} defined two functions called interpretation-functions: $\mbox{DEK}$ and $\mbox{DEKAN}$. These functions safely evaluate the level of security of any atomic message. They operate on any message either it contains variables or it is a ground term. The variables are treated the same way as the ground atoms. The function $\mbox{DEK}$ selects the direct encryption key of an atomic message $\alpha$, that returns the identity of agents that know the reverse form of that key from the context.  The function $\mbox{DEKAN}$ returns in addition to the identity of agents that know the reverse form of the direct key in the context, all the neighbors of $\alpha$ encrypted with that key. For example, we have:
\begin{itemize}
\item $\mbox{DEK}(\alpha,\{\alpha.C.X\}_{k_{\textcolor{blue}{ab}}})= \ulcorner k_{ab}^{-1}\urcorner= \{\textcolor{blue}{A},\textcolor{blue}{B}\}$
\item $\mbox{DEKAN}(\alpha,\{\alpha.\textcolor{blue}{C}.\textcolor{blue}{X}\}_{k_{\textcolor{blue}{ab}}})=\{C\}\cup \{\overline{X}\}\cup \ulcorner k_{ab}^{-1}\urcorner=\{C\}\cup \{\overline{X}\}\cup \{A,B\}=\{\textcolor{blue}{C}, \textcolor{blue}{\overline{X}}, \textcolor{blue}{A}, \textcolor{blue}{B}\}$
\end{itemize}
Where $X$ is a variable. The notation $\overline{X}$ refers to the set of agent identities in $X$ after being substituted, which is only known at runtime. Please notice that the function $\mbox{DEKAN}$ is not variable free (i.e. ${X}$ is among the returned values). Now that we have these two functions, it is possible to evaluate the security level of any atom in any message. Hence, we can compare the level of security levels on sent messages with those in received messages and check whether a protocol is increasing or not. Theorem \ref{mainTh} establishes its correctness if it is shown increasing.

\section{Witness-Functions}\label{sectionWF}
The witness-functions have been proposed by Fattahi et al in ~\cite{TheseJF,ChapterJFSpringer,WitnessArt1,Relaxed}. First, we define six functions $F_{MAX}^{IK}, F_{IK}^{IK}, F_{N}^{IK}, F_{MAX}^{EK}, F_{EK}^{EK},$ and $F_{N}^{EK}$ and we prove that they are safe functions. For lack of space, we will focus on the function $F_{MAX}^{IK}$ only and we refer to it as $F$. This function $F$ operates on ground terms only and does not deal with variables. To rank the security level of an atom $\alpha$  in a message $m$, this function selects the reverse key of the innermost protective key, and returns the identity of agents that are allowed to know this key in the context, in addition to all the neighbor identities of $\alpha$ in $m$ encrypted with this key. The innermost protective key is not necessarily the direct key but the most internal encryption key that has a security level superior that the security level of $\alpha$ given in the context. This function is not very useful in practice because a static analysis should take in consideration variables. To deal with variables, we use the derivative form of $F$ instead of $F$. \\
\begin{defini}{(Derivative Function)}\label{Fder}  $ $\\
\footnotesize{
\[
F'(\alpha,m\sigma)=F(\alpha, \partial [{\alpha}] m\sigma) = \left\{
\begin{array}{ll}
F(\alpha,\partial m) & \mbox{if } \alpha \in {\mathcal{A}}(\partial m),\\
F(X,\partial [{X}] m) & \mbox{if }\alpha \notin {\mathcal{A}}(\partial m) \\
& \mbox{and } \alpha =X\sigma, \forall \sigma
\end{array}
\right.
\]
}
\end{defini}

Since $F'(\alpha,m\sigma)$ does not depend on substitution (i.e the run $\sigma$), we denote it simply by $F'(\alpha,m)$. Although the derivative function $F'$ eliminates the effect of variables, it is not yet good enough to analyze protocols. For example, for a valid trace $m=\{\alpha.A.B\}_{k_{cd}}­$ having two sources $m_1=\{\alpha.A.X\}_{k_{cd}}­$ and $m_2=\{\alpha.Y.B\}_{k_{cd}}­$ ($X$ and $Y$ are variables), it may return two different images (two security levels). For instance,
\begin{itemize}
\item $F'(\alpha,\{\alpha.A.X\}_{k_{cd}})=F(\alpha,\{\alpha.\textcolor{blue}{A}.\bcancel{\textcolor{red}{X}}\}_{k_{\textcolor{blue}{cd}}})=F(\alpha,\{\alpha.\textcolor{blue}{A}\}_{k_{\textcolor{blue}{cd}}})=\{A, C, D\}$
\item $F'(\alpha,\{\alpha.Y.B\}_{k_{cd}})=F(\alpha,\{\alpha.\bcancel{\textcolor{red}{Y}}.\textcolor{blue}{B}\}_{k_{\textcolor{blue}{cd}}})=F(\alpha,\{\alpha.\textcolor{blue}{B}\}_{k_{\textcolor{blue}{cd}}})=\{B, C, D\}$
\end{itemize}
For that, we define the witness-function. A witness function takes $F$ and the protocol $p$ as parameters, then looks for  all the sources of a ground term $m\sigma$ in the finite set  ${\mathcal{M}}_p^{\mathcal{G}}$, then applies $F'$ to all of them, and finally returns the minimum. This minimum is obviously unique. 
\begin{defini}{[Witness-Function]}\label{WF}
\[{{{\mathcal{W}}}}_{p,F}(\alpha,m\sigma)=\underset{{\{(m',\sigma') \in \tilde{\mathcal{M}}_p^{\mathcal{G}}\otimes\Gamma|m'\sigma' = m \sigma \}}}{\sqcap} \!\!\!\!\!\!\!\!\!\!\!\!\!\!\!\!\!\!\!\!\!\!\!\!\!\!\!F'(\alpha, m'\sigma')\]
\end{defini}


Using a witness-function roughly is not realistic since we cannot predict all the valid traces $m\sigma$ and their sources in the protocol statically. For that, we bind witness-function into two bounds that do not depend on substitution (i.e on $\sigma$). The upper bound is $ F'(\alpha, m)$ and returns the smallest set of principal identities for any $\alpha$ in $m$ whereas the lower bound, which is $\underset{{\{(m',\sigma') \in \tilde{\mathcal{M}}_p^{\mathcal{G}}\otimes\Gamma|m'\sigma' =m \sigma' \}}}{\sqcap} \!\!\!\!\!\!\!\!\!\!\!\!\!\!\!\!\!\!\!\!\!\!\!\!\!\!\!\!\!F'(\alpha,  m'\sigma') $), returns the largest set of identities from all the possible sources of $m$ in the protocol (the messages that are unifiable with $m$). Considering these facts and Theorem \ref{mainTh}, the theorem of secrecy decision with a witness-function becomes as follows.\\
\begin{thm}{[Decision for Secrecy]}\label{PAT}$ $\\
$p$ is correct with respect to secrecy if:
$\forall R.r, \forall \alpha \in {\mathcal{A}}{(r^+ )}$, we have:
$$\underset{{\{(m',\sigma') \in \tilde{\mathcal{M}}_p^{\mathcal{G}}\otimes\Gamma|m'\sigma' = r^+ \sigma' \}}}{\sqcap} \!\!\!\!\!\!\!\!\!\!\!\!\!\!\!\!\!\!\!\!\!\!\!\!\!\!\!\!\!F'(\alpha,  m'\sigma') \sqsupseteq \ulcorner \alpha \urcorner \sqcap F'(\alpha, R^-)$$
\end{thm}

This theorem is the one used to prove that a protocol is increasing or not. Its correctness for secrecy  follows. Please note that when we say that we run an analysis with a witness-function, we rather mean an analysis using the two bounds of the witness-function, not with the witness-function itself.

%% file: someProtocolSMC2017WFvsIF.tex
\section{Protocol Definition and Context Setup}  \label{sectionAnWL2}

We analyze the protocol defined by $p$ in Table~\ref{WLMV:protv2p} for secrecy. The aim of our analysis is to show that $p$ does not divulge the secret $sec$, supposed to be shared between $A$ and $S$ only. We show that the interpretation-functions fail, whereas a witness-function succeeds.

\begin{center}
          \begin{table}[h]
      \begin{center}    
         \caption{Protocol Definition}
         \label{WLMV:protv2p} 
               $\begin{array}{|l|}
               \hline\\
                   \begin{array}{llll}
                    p   ::&\langle 1,A\rightarrow S: \{A.N_a.S.B\}_{k_s}\rangle.  \\
                    & \langle 2,S\rightarrow B: \{B.A.S.N_a\}_{k_b}.\{A.B.S.\{S.sec\}_{k_a}\}_{k_b}\rangle. \\
                    & \langle 3,B\rightarrow A:\{B.\{S.sec\}_{k_a}.A.N_a.S\}_{k_{a}}\rangle\\ &\\
                    \end{array} \\ \hline  \end{array}$
\end{center}
         \end{table}
\end{center}

    The   role-based   specification  of $p$  is ${\cal R}_G(p) = \{{\cal A}_G ,~{\cal B}_G ,~ {\cal S}_G \}$,
    where the generalized role ${\cal A}_G $ of $A$ is as follows:
\footnotesize{
    \[\begin{array}{l}
             \begin{array}{llllll}
                    {\cal A}_G =& \langle  i.1,  A    & \rightarrow & I(S):&  \{A.N_a^{i}.S.B\}_{k_s} \rangle.\\
                    & \langle i.2,  I(B)    & \rightarrow & A:&  \{B.\{S.X\}_{k_a}.A.N_a^{i}.S\}_{k_{a}}\rangle
             \end{array}\end{array}\]
}
\normalsize
     The generalized role ${\cal B}_G $  of $B$ is as follows:
\footnotesize{
            \[\begin{array}{l}
            \begin{array}{lllll}
                {\cal B}_G =& \langle i.1,  I(S) & \rightarrow & B:& \{B.A.S.Y\}_{k_b}.\{A.B.S.Z\}_{k_b} \rangle.\\
                        & \langle i.2,  B    & \rightarrow & I(A) :&  \{B.Z.A.Y.S\}_{k_{a}} \rangle  \\
            \end{array}
             \end{array}\]
}
\normalsize
             The  generalized role $ {\cal S}_G$ of $S$ is as follows:
\footnotesize{
            \[\begin{array}{l}
             \begin{array}{lllll}
                            {\cal S}_G = & \langle i.1, I(A) & \rightarrow & S:& \{A.T.S.B\}_{k_s}
                             \rangle. \\
                                    & \langle i.2,S & \rightarrow & I(B):&\{B.A.S.T\}_{k_b}.\{A.B.S.\{S.sec\}_{k_a}\}_{k_b}\rangle\\
               \end{array}
            \end{array} \]
}     
\normalsize       
Let us have a context of verification such that: \\

$\ulcorner k_{a}\urcorner=\ulcorner k_{b}\urcorner = \ulcorner k_{s}\urcorner=\bot$ (public keys); 

$\ulcorner k_{a}^{-1}\urcorner = \{A\}$; $\ulcorner k_{b}^{-1}\urcorner = \{B\}$; $\ulcorner k_{s}^{-1}\urcorner = \{S\}$ (private keys);

$\ulcorner N_a^i \urcorner=\{A, B, S\}$ (nonce shared between $A, B$ and $S$); 

$\ulcorner sec \urcorner=\{A, S\}$ (secret shared between $A$ ans $S$ \textbf{\underline{only}}); 

$\forall \mbox{ agID } \in {\cal{I}}, \ulcorner \mbox{ agID } \urcorner=\bot$ (agents' identities are public);

The variables are denoted by $X$, $Y$, $Z$ and $T$ in the generalized roles.\\



\section{Protocol analysis with the Interpretation-Functions} \label{sectionAnWL222}

For the sake of conciseness, we do not analyze the entire protocol. We analyze just the generalized role of $B$ where the Interpretation-Functions fail to show the growth of the protocol. In that generalized role, $B$ receives $R^{-}=\{A.S.Y\}_{k_b}.\{A.B.S.Z\}_{k_b}$ and sends $r^{+}=\{B.Z.A.Y\}_{k_{a}}$. 

\subsection{Analysis with the DEK function} \label{sectionAnWL5486}
\begin{enumerate}

\item Receiving step: $\mbox{DEK}(Y,R^{-}) = \mbox{DEK}(Y, \{B.A.S.Y\}_{k_b}.\{A.B.S.Z\}_{k_b})=$ $\mbox{DEK}(Y, \{B.A.S.Y\}_{k_b}) \sqcap \mbox{DEK}(Y, \{A.B.S.Z\}_{k_b})=$ $\ulcorner k_b^{-1} \urcorner \sqcap \top=\ulcorner k_b^{-1} \urcorner=\{B\}$
\item Sending step: $\mbox{DEK}(Y,r^{+}) = \mbox{DEK}(Y, \{B.Z.A.Y.S\}_{k_{a}})=\ulcorner k_a^{-1} \urcorner=\{A\}$
\end{enumerate}
As we can see, we cannot obtain the result $\mbox{DEK}(Y,r^{+})  \sqsupseteq \ulcorner Y \urcorner \sqcap \mbox{DEK}(Y,R^{-})$. So, the DEK function \textcolor{red}{fails}.

\subsection{Analysis with the \mbox{DEKAN} function} \label{sectionAnWL877}

\begin{enumerate}

\item On receiving: $\mbox{DEKAN}(Y,R^{-}) = \mbox{DEKAN}(Y, \{B.A.S.Y\}_{k_b}.\{A.B.S.Z\}_{k_b})=$ $\mbox{DEKAN}(Y, \{B.A.S.Y\}_{k_b}) \sqcap \mbox{DEKAN}(Y, \{A.B.S.Z\}_{k_b})=$ $\ulcorner k_b^{-1} \urcorner \sqcap \{ B, A, S\} \sqcap \top=\ulcorner k_b^{-1} \urcorner  \sqcap \{B, A, S\} =\{B\} \cup \{B, A, S\}=  \{A, B, S\}$
\item On sending: $\mbox{DEKAN}(Y,r^{+}) = \mbox{DEKAN}(Y, \{B.Z.A.Y.S\}_{k_{a}})=\ulcorner k_a^{-1} \urcorner \cup \{B, \textcolor{red}{\overline{Z}}, A, S\}=\{B, \textcolor{red}{\overline{Z}}, A, S\}$
\end{enumerate}
As we can see, we cannot obtain the result $\mbox{DEKAN}(Y,r^{+})  \sqsupseteq \ulcorner Y \urcorner \sqcap \mbox{DEKAN}(Y,R^{-})$. So, the DEKAN function \textcolor{red}{fails}, as well.

\section{Protocol analysis with the witness-unctions} \label{sectionAnWL222333}

Let $F= F_{MAX}^{IK}$; ${\mathcal{W}}_{p,F}= {\mathcal{W}}_{{p},F_{MAX}^{IK}}$;\\

We denote  by ${\Upsilon}_{p,F}(\alpha,m)$ the lower bound $\underset{\{{(m',\sigma') \in \tilde{\cal{M}}_{p}^{\cal{G}}}\otimes \Gamma{|m'\sigma' = m\sigma' \}}}{\sqcap}\!\!\!\!\!\!\!\!\!\!  F'(\alpha, m'\sigma')$ of the witness-function ${\mathcal{W}}_{p,F}(\alpha,m)$.\\

The set of encryption patterns generated by $p$ is  $\tilde{\cal{M}}_{p}^{\cal{G}}=\{
 \{A_1.N_{A_1}.S_1.B_1\}_{K_{S_1}},
\{B_2.\{S_2.X_2\}_{K_{A_2}}.A_2.N_{A_2}.S_2\}_{K_{A_2}},\\
\{B_3.A_3.S_3.Y_3\}_{K_{B_3}}, \{A_4.B_4.S_4.Z_4\}_{K_{B_4}},\\
 \{B_5.Z_5.A_5.Y_5.S_5\}_{K_{A_5}},\\ \{A_6.T_6.S_6.B_6\}_{K_{S_6}}, 
 \{B_7, A_7.S_7.T_7\}_{K_{B_7}},\\ \{A_8.B_8.S_8.\{S_8.sec_8\}_{K_{A_8}}\}_{K_{B_8}}
 \}$\\
 
 The renamed variables in $\tilde{\cal{M}}_{p}^{\cal{G}}$ are denoted by $X_2, Y_3, Z_4, Z_5, Y_5, T_6$ and $T_7$;

\subsection{Analysis of the Generalized Roles of $A$}

According to the generalized role of  $A$, an agent $A$ may take part in some session $S^{i}$  in which he receives nothing (i.e. $\epsilon$) and sends the message $\{A.N_a^{i}.S.B\}_{k_s}$. This is described by the following rule: \[{S^{i}}:\frac{\epsilon}{ \{A.N_a^{i}.S.B\}_{k_s}}\]

\noindent{-Analysis of the messages exchanged  in $S^{i}$:}\\

1- For $N_a^{i}$:\\

a- Receiving step: $R_{S^{i}}^-=\epsilon$ \textit{(when receiving, we use the upper bound)}

\begin{equation}
F'(N_a^{i}, R_{S^{i}}^-)=F(N_a^{i},\partial  [{N_a^{i}}]\epsilon)=F(N_a^{i},\epsilon)=\top
\label{eq1}
\end{equation}

b- Sending step: $r_{S^{i}}^+= \{A.N_a^{i}.S.B\}_{k_s}$ \textit{(when sending , we use the lower bound)}\\

Since $\{A.N_a^{i}.S.B\}_{k_s}$ is a ground term (no variable in), then we have: ${\Upsilon}_{p,F}(N_a^{i}, \{A.N_a^{i}.S.B\}_{k_s})= $ $F(N_a^{i}, \{A.N_a^{i}.S.B\}_{k_s})$\\

$\mbox{Since } F=F_{MAX}^{IK}$, we have:\\

$F(N_a^{i}, \{A.N_a^{i}.S.B\}_{k_s})=\ulcorner k_{s}^{-1} \urcorner \cup \{A, S, B\}$. Then:\\

\begin{equation}
{\Upsilon}_{p,F}(N_a^{i}, \{A.N_a^{i}.S.B\}_{k_s})=\{A, S, B\} \label{eq2}
\end{equation}
\\
2- Conformity  with Theorem \ref{PAT}:\\ 

From (\ref{eq1}) and (\ref{eq2}) and since $ \ulcorner N_a^{i} \urcorner=\{A, B, S\}$ in the context, we have:
\begin{equation}
{\Upsilon}_{p,F}(N_a^{i}, r_{S^{i}}^+) \sqsupseteq \ulcorner N_a^{i} \urcorner \sqcap F'(N_a^{i}, R_{S^{i}}^-)
\label{eq3}
\end{equation}

Then,  the generalized role of $A$ respects Theorem \ref{PAT}. ~~~~ (I)

\subsection{Analysis of the generalized roles of $B$}
According to the generalized role of $B$, an agent $B$ participates in a session $S^{i}$ in which he receives the message $\{B.A.S.Y\}_{k_b}.\{A.B.S.Z\}_{k_b}$ and sends the message $\{B.Z.A.Y.S\}_{k_{a}}$. This is described by the following rule:

\[{S^{i}}:\frac{\{B.A.S.Y\}_{k_b}.\{A.B.S.Z\}_{k_b}}{\{B.Z.A.Y.S\}_{k_{a}}}\]

1-$\forall Z$:\\

a- Receiving step: $R_{S^{i}}^-=\{B.A.S.Y\}_{k_b}.\{A.B.S.Z\}_{k_b}$ \textit{(when receiving, we use the upper bound)}
\footnotesize{
\begin{equation} 
\begin{tabular}{lll}
   $F'(Z,R_{S^{i}}^-)$  &$=$& $F'(Z,\{B.A.S.Y\}_{k_b}.\{A.B.S.Z\}_{k_b} )$\\
    & $=$ & $F(Z,\partial  [{Z}]\{B.A.S.Y\}_{k_b}.\{A.B.S.Z\}_{k_b})$  \\
   & $=$ & $F(Z,\{B.A.S\}_{k_b}.\{A.B.S.Z\}_{k_b} )$ \\
   & $=$ & $ F(Z,\{B.A.S\}_{k_b}) \sqcap F(Z,\{A.B.S.Z\}_{k_b} )  $ \\
    & $=$  & $\top \sqcap  F(Z,\{A.B.S.Z\}_{k_b} ) $ \\
    & $=$  & $ F(Z,\{A.B.S.Z\}_{k_b} ) $ \\
    & $=$  & $\ulcorner k_{b}^{-1} \urcorner \cup \{A,B, S\} $ \\
    & $=$  & $\{A, B, S\}$ 
\end{tabular}
\label{eq4}
\end{equation}
}
\normalsize

b- Sending step: $r_{S^{i}}^+=\{B.Z.A.Y.S\}_{k_{a}}$ \textit{(when sending , we use the lower bound)}\\

 $\forall Z.\{(m',\sigma') \in \tilde{\cal{M}}_{p}^{\cal{G}}\otimes\Gamma|{m'\sigma' = r_{S^{i}}^+\sigma' } \}$\\
 
 $=\forall Z.\{(m',\sigma') \in \tilde{\cal{M}}_{p}^{\cal{G}}\otimes\Gamma|{m'\sigma' = \{B.Z.A.Y.S\}_{k_{a}} \sigma' } \}$ \\
 
 $=\{(  \{B_5.Z_5.A_5.Y_5.S_5\}_{K_{A_5}},\sigma_1')\}$ such that: 
 
 $\sigma_1'=\{ B_5 \longmapsto B, Z_5\longmapsto Z, A_5 \longmapsto A, Y_5 \longmapsto Y,$\\  $~~~~~~~~~~~~~S_5 \longmapsto S, K_{A_{5}}\longmapsto  k_{a}\}$\\

${\Upsilon}_{p,F}(Z,\{B.Z.A.Y.S\}_{k_{a}})$\\

$=\{\mbox{Definition of the lower bound of the witness-function}\}$\\

$F'(Z,\{B_5.Z_5.A_5.Y_5.S_5\}_{K_{A_5}} \sigma_{1}') $\\

$=\{\mbox{Setting the static neighborhood}\}$\\

$F'(Z, \{B.Z_5.A.Y.S\}_{k_{a}} \sigma_{1}')$ \\

$=\{\mbox{Definition } \ref{Fder}\}$\\

$F(Z_5,\partial[{Z_5}]  \{B.Z_5.A.Y.S\}_{k_{a}})$ \\

$=\{\mbox{Derivation}\}$
\\

$F(Z_5, \{B.Z_5.A.S\}_{k_{a}})$ \\

$=\{\mbox{Since } F=F_{MAX}^{IK}\}$\\

$\ulcorner k_{a}^{-1} \urcorner \cup \{B, A, S\}=\{A, B, S\}$\\

Then, we have:

\begin{equation}
{\Upsilon}_{p,F}(Z,\{B.Z.A.Y.S\}_{k_{a}})=\{A, B, S\}
\label{eq5}
\end{equation}

2- $\forall Y$:\\

a- Receiving step: $R_{S^{i}}^-=\{A.S.Y\}_{k_b}.\{A.B.S.Z\}_{k_b}$ \textit{(when receiving, we use the upper bound)}
\footnotesize{
\begin{equation} 
\begin{tabular}{lll}
   $F'(Y,R_{S^{i}}^-)$  &$=$& $F'(Y,\{B.A.S.Y\}_{k_b}.\{A.B.S.Z\}_{k_b} )$\\
    & $=$ & $F(Y,\partial  [{Y}]\{B.A.S.Y\}_{k_b}.\{A.B.S.Z\}_{k_b})$  \\
   & $=$ & $F(Y,\{B.A.S.Y\}_{k_b}.\{A.B.S\}_{k_b} )$ \\
   & $=$ & $ F(Y,\{B.A.S.Y\}_{k_b}) \sqcap F(Y,\{A.B.S\}_{k_b} )  $ \\
   & $=$ & $ F(Y,\{B.A.S.Y\}_{k_b}) \sqcap\top  $ \\
   & $=$ & $ F(Y,\{B.A.S.Y\}_{k_b})$ \\
    & $=$  & $\ulcorner k_{b}^{-1} \urcorner \cup \{B, A, S\} $ \\
    & $=$  & $\{A, B, S\}$ \\
\end{tabular}
\label{eq6}
\end{equation}
}
\normalsize

b- Sending step: $r_{S^{i}}^+=\{B.Z.A.Y.S\}_{k_{a}}$ \textit{(when sending , we use the lower bound)}\\

 $\forall Y.\{(m',\sigma') \in \tilde{\cal{M}}_{p}^{\cal{G}}\otimes\Gamma|{m'\sigma' = r_{S^{i}}^+\sigma' } \}$\\
 
 $=\forall Y.\{(m',\sigma') \in \tilde{\cal{M}}_{p}^{\cal{G}}\otimes\Gamma|{m'\sigma' = \{B.Z.A.Y.S\}_{k_{a}} \sigma' } \}$ \\
 
 $=\{(  \{B_5.Z_5.A_5.Y_5.S_5\}_{K_{A_5}},\sigma_1')\}$ such that: 
 
 $\sigma_1'=\{ B_5 \longmapsto B, Z_5\longmapsto Z, A_5 \longmapsto A, Y_5 \longmapsto Y,$\\  $~~~~~~~~~~~~~S_5 \longmapsto S, K_{A_{5}}\longmapsto  k_{a}\}$\\

${\Upsilon}_{p,F}(Y, \{B.Z.A.Y.S\}_{k_{a}})$\\

$=\{\mbox{Definition of the lower bound of the witness-function}\}$\\

$F'(Y, \{B_5.Z_5.A_5.Y_5.S_5\}_{K_{A_5}} \sigma_{1}') $\\

$=\{\mbox{Setting the static neighborhood}\}$\\

$F'(Y, \{B.Z.A.Y_5.S\}_{k_{a}} \sigma_{1}')$ \\

$=\{\mbox{Definition } \ref{Fder}\}$\\

$F(Y_5,\partial[{Y_5}]  \{B.Z.A.Y_5.S\}_{k_{a}})$ \\

$=\{\mbox{Derivation} \}$
\\

$F(Y_5, \{B.A.Y_5.S\}_{k_{a}})$ \\

$=\{\mbox{Since } F=F_{MAX}^{IK}\}$\\

$\ulcorner k_{a}^{-1} \urcorner \cup \{B, A, S\}=\{A, B, S\}$\\

Then, we have:

\begin{equation}
{\Upsilon}_{p,F}(Y,\{B.Z.A.Y.S\}_{k_{a}})=\{A, B\}
\label{eq7}
\end{equation}

3- Conformity  with Theorem \ref{PAT}:\\

From (\ref{eq4}) and (\ref{eq5}), we have:
\begin{equation}
{\Upsilon}_{p,F}(Z, r_{S^{i}}^+) \sqsupseteq \ulcorner Z \urcorner \sqcap F'(Z, R_{S^{i}}^-)
\label{eq8}
\end{equation}

From (\ref{eq6}) and (\ref{eq7}), we have:
\begin{equation}
{\Upsilon}_{p,F}(Y, r_{S^{i}}^+) \sqsupseteq \ulcorner Y \urcorner \sqcap F'(Y, R_{S^{i}}^-)
\label{eq9}
\end{equation}

From (\ref{eq8}) and (\ref{eq9}), we have:  the generalized role of $B$ respects Theorem \ref{PAT}. ~~~~~~~~~~~~~~~~~~~~~~~~~~~~~~~~~~~~~~~~~~~~~~~~~~~~~(II)\\

\subsection{Analysis of the generalized roles of $S$}

According to the generalized role of $B$, an agent $B$ participates in a session $S^{i}$ in which he receives the message $\{A.T.S.B\}_{k_s}$ and sends the message $\{B.A.S.T\}_{k_b}.\{A.B.S.\{S.sec\}_{k_a}\}_{k_b}$. This is described by the following rule:

\[{S^{i}}:\frac{\{A.T.S.B\}_{k_s}}   {\{B.A.S.T\}_{k_b}.\{A.B.S.\{S.sec\}_{k_a}\}_{k_b}}\]

1- $\forall T$:\\

a- Receiving step: $R_{S^{i}}^-=\{A.T.S.B\}_{k_s}$ \textit{(when receiving, we use the upper bound)}
\footnotesize{
\begin{equation} 
\begin{tabular}{lll}
   $F'(T,\{A.T.S.B\}_{k_s})$ & $=$ & $F(T,\partial  [{T}]\{A.T.S.B\}_{k_s})$  \\
   & $=$ & $ F(T,\{A.T.S.B\}_{k_s} )$ \\
    & $=$  & $\ulcorner k_{s}^{-1} \urcorner \cup \{A, S, B\} $ \\
    & $=$  & $\{A, B, S\}$ \\
\end{tabular}
\label{eq10}
\end{equation}
}
\normalsize
b- Sending step: $r_{S^{i}}^+=\{B.A.S.T\}_{k_b}.\{A.B.S.\{S.sec\}_{k_a}\}_{k_b}$ \textit{(when sending , we use the lower bound)}. We have: 
\footnotesize{
\begin{equation}
\begin{tabular}{lll}
   ${\Upsilon}_{p,F}(T,r_{S^{i}}^+)$ & $=$ & ${\Upsilon}_{p,F}(T,\{B.A.S.T\}_{k_b}.\{A.B.S.\{S.sec\}_{k_a}\}_{k_b})$  \\
   & $=$ & ${\Upsilon}_{p,F}(T,\{B.A.S.T\}_{k_b}) \sqcap$ \\
   & $ $ & ${\Upsilon}_{p,F}(T,\{A.B.S.\{S.sec\}_{k_a}\}_{k_b}) $ \\
   & $=$ & ${\Upsilon}_{p,F}(T,\{B.A.S.T\}_{k_b}) \sqcap \top$ \\
   & $=$ & ${\Upsilon}_{p,F}(T,\{B.A.S.T\}_{k_b})$ 
\end{tabular}
\label{eq11}
\end{equation}
}
\normalsize
 $\forall T.\{(m',\sigma') \in \tilde{\cal{M}}_{p}^{\cal{G}}\otimes\Gamma|{m'\sigma' = r_{S^{i}}^+\sigma' } \}$\\
 
 $=\forall T.\{(m',\sigma') \in \tilde{\cal{M}}_{p}^{\cal{G}}\otimes\Gamma|{m'\sigma' = \{B.A.S.T\}_{k_b}\}_{k_{bs}} \sigma' } \}$ \\
 
 $=\{(\{A_7.S_3.Y_1\}_{K_{B_2}},\sigma_1')\}$ such that: 
 
 $$\sigma_1'=\{ B_7 \longmapsto B, A_7 \longmapsto A, S_7 \longmapsto S, T_7 \longmapsto T, K_{B_{7}}\longmapsto  k_{b}\}$$

${\Upsilon}_{p,F}(T,\{B.A.S.T\}_{k_b})$\\

$=\{\mbox{Definition of the lower bound of the witness-function}\}$\\

$F'(T,\{B_7.A_7.S_7.T_7\}_{K_{B_4}}\sigma_{1}') $\\

$=\{\mbox{Setting the static neighborhood}\}$\\

$F'(T, \{B.A.S.T_7\}_{k_{b}} \sigma_{1}')$ \\

$=\{\mbox{Definition } \ref{Fder}\}$
\\

$F(T_7,\partial[{T_7}] \{B.A.S.T_7\}_{k_{b}})$ \\

$=\{\mbox{Derivation} \}$
\\

$F(T_7,\{B.A.S.T_7\}_{k_{b}})$ \\

$=\{\mbox{Since } F=F_{MAX}^{IK}\}$\\

$\ulcorner k_{b}^{-1} \urcorner \cup \{B, A, S\}=\{A, B, S\}$\\

Then, we have:

\begin{equation}
{\Upsilon}_{p,F}(T,\{B.A.S.T\}_{k_b})=\{A, B, S\}
\label{eq12}
\end{equation}

2- For the secret $sec$:\\

a- Receiving step: $R_{S^{i}}^-=\{A.T.S.B\}_{k_s}$ \textit{(when receiving, we use the upper bound)}
\footnotesize{
\begin{equation} 
\begin{tabular}{lll}
   $F'(sec,\{A.T.S.B\}_{k_s})$ & $=$ & $F(sec,\{A.S.B\}_{k_s})$  \\
   & $=$ & $\top$ \\
\end{tabular}
\label{eq13}
\end{equation}
}
\normalsize
b- Sending step: $r_{S^{i}}^+=\{B.A.S.T\}_{k_b}.\{A.B.S.\{S.sec\}_{k_a}\}_{k_b}$ \textit{(when sending , we use the lower bound)}
\footnotesize{
\begin{equation}
\begin{tabular}{lll}
   ${\Upsilon}_{p,F}(sec,r_{S^{i}}^+)$ & $=$ & ${\Upsilon}_{p,F}(sec,\{B.A.S.T\}_{k_b}.$  \\
    & & $\{A.B.S.\{S.sec\}_{k_a}\}_{k_b})$  \\
   & $=$ & ${\Upsilon}_{p,F}(sec,\{B.A.S.T\}_{k_b}) \sqcap$ \\
   & $ $ & ${\Upsilon}_{p,F}(sec,\{A.B.S.\{S.sec\}_{k_a}\}_{k_b}) $ \\
   & $=$ & $\top \sqcap {\Upsilon}_{p,F}(sec,\{A.B.S.\{S.sec\}_{k_a}\}_{k_b})$ \\
   & $=$ & ${\Upsilon}_{p,F}(sec,\{A.B.S.\{S.sec\}_{k_a}\}_{k_b})$ 
\end{tabular}
\label{eq14}
\end{equation}
}
\normalsize
Since $\{A.B.S.\{S.sec\}_{k_a}\}_{k_b})$ is a ground term, we have directly:
\footnotesize{
\begin{equation}
\begin{tabular}{lll}
   ${\Upsilon}_{p,F}(sec,\{A.B.S.\{S.sec\}_{k_a}\}_{k_b})$ & $=$ &   \\
   $F(sec,\{A.B.S.\{S.sec\}_{k_a}\}_{k_b})$ & $=$ & \\ 
     $~~~~~~~~~~~~~~~~~~~~~~\{\mbox{Since } F=F_{MAX}^{IK}\}$ & &\\ 
  $\ulcorner {k_a}^{-1}\urcorner \cup \{S\} $ & $=$ & $\{A, S\}$ 
\end{tabular} 
\label{eq15}
\end{equation}
}
\normalsize
3- Conformity  with Theorem \ref{PAT}:\\
$ $\\
From (\ref{eq10}) and (\ref{eq12}), we have:
\footnotesize{
\begin{equation}
{\Upsilon}_{p,F}(T,r_{S^{i}}^+) \sqsupseteq \ulcorner T \urcorner \sqcap F'(T,R_{S^{i}}^- )
\label{eq16}
\end{equation}
}
\normalsize
From (\ref{eq13}) and (\ref{eq15}), we have:
\footnotesize{
\begin{equation}
{\Upsilon}_{p,F}(sec,r_{S^{i}}^+) \sqsupseteq \ulcorner sec \urcorner \sqcap F'(sec,R_{S^{i}}^- )
\label{eq17}
\end{equation}
\normalsize
From (\ref{eq16})  and (\ref{eq17}),  we have:  the generalized role of $S$ respects Theorem \ref{PAT}. ~~~~~~~~~~~~~~~~~~~~~~~~~~~~~~~~~~~~~~~~~~~~~~~~(III)$ $\\
From (I) and (II) and (III), we conclude that: $p$ respects Theorem~\ref{PAT}. So it is increasing, then, it is correct for secrecy. ~~~~(IV)